\documentclass[12pt]{iopart}

\usepackage{graphicx}
\usepackage{epstopdf}
\usepackage{subfigure}
\usepackage{bm}
\usepackage{cite}
\begin{document}

\title{Proposals for gain cascading in single-pass of a free-electron laser oscillator}

\author{Kai Li$^{1,2}$ and Haixiao Deng$^1$}

\address{$^1$Shanghai Institute of Applied Physics, Chinese Academy of Sciences, Shanghai 201800, China\\
 $^2$University of Chinese Academy of Sciences, Beijing 100049, China.}
\ead{denghaixiao@sinap.ac.cn}
\vspace{10pt}
\begin{indented}
\item[]PACS numbers: 41.60.Cr
\end{indented}

\begin{abstract}
The low-gain free-electron laser (FEL) oscillators are cutting-edge tools to produce fully coherent radiation in the spectral region from terahertz to vacuum ultraviolet, and potentially in hard X-ray. In this paper, it is proposed to utilize an FEL oscillator with multi-stage undulators to enable gain cascading in a single-pass, making it possible to achieve shorter single pulse lengths, higher peak power, and even higher pulse energy than classical FEL oscillator. Theoretical analysis and numerical simulations in the infrared and hard X-ray regions show that our proposal is effective.
\end{abstract}

%
%
%
%
%

\section{Introduction}

While the lasing wavelengths of conventional lasers are limited by the energy states of gain media, a free-electron laser (FEL), utilizing the electron beam as a broadband gain medium, is capable of delivering electromagnetic radiation ranging from the infrared to hard X-ray regions, depending on the electron beam energy and magnetic field strength of the undulator \cite{1pc}. In the early stage, most of FELs operated as multi-pass low-gain oscillators, producing low-energy photons \cite{2elr,3vapw,4ngr,5wyk,6gng}. More recently, the successful operations of single-pass high-gain FEL user facilities \cite{7aw,8ep,9it,10ae,10br,10cbr} in the soft and hard x-ray regimes enable the simultaneous probing of both the ultra-small and ultra-fast worlds. The high brightness and wavelength tunability make the FEL a unique and innovative tool for the investigation of matter. Therefore, a number of FEL facilities, including both the oscillators and amplifiers, are under construction and consideration all over the world.

There is, however, a strong scientific demand to achieve even higher pulse powers and/or shorter pulse lengths for FEL applications such as bio-imaging, nonlinear optics and time-resolved experiments \cite{11fa,12ss}. On one hand, many ideas have been proposed to further shorten the FEL pulse length, either by employing external lasers (see, for instance, Refs. \cite{13sel,14zaa,15sel,16xd,17yj}), or by producing ultra-short electron bunches \cite{18ep,19rs}, or by combining these two approaches as mode locking FEL \cite{19atnr,19bddj,19cmbw}; on the other hand, the FEL peak power enhancement usually comes down to the technique of tapered undulator \cite{20knm}. Further, FEL schemes that produce higher power and shorter pulses simultaneously are also intensively investigated \cite{21tt,22pe}. These schemes, however, are tailored for single-pass FEL amplifiers, and cannot be easily applied to multi-pass FEL oscillators with low single-pass gain, round-trip growth, and saturation due to dynamic equilibrium.

Besides, the FEL community is continuing to develop more sophisticated schemes in pursuit of e.g. full coherence \cite{23ylh,24sg,25dh,26aj} and multi-color operation \cite{27ht,28laa,29ma}, in which the techniques of fresh bunch and multi-bunch are frequently used. Inspired by these earlier works and the many applications, in this paper, a new configuration of FEL oscillator which generates shorter pulse duration and higher peak power is proposed. Although the impacts of gain cascading and longer undulators are obvious in single-pass FEL, its effect to FEL oscillator has not been investigated systematically. A brief theoretical analysis is presented in the second section, which shows the principles of the new scheme. In the third and forth part, numerical simulations of 1.6 $\mu $m infrared and 1 $\rm{\AA}$ X-ray FEL oscillator (XFELO) are studied and the results demonstrate that, by using a gain cascading scheme in single-pass, FEL pulses can be delivered with several-fold of magnitude higher power and pulse energy with shorter temporal duration compared to conventional FEL oscillators. At last, a summary of this paper is presented and the future research work is discussed.

\section{Theoretical analysis}
The principle of an FEL oscillator is well-known. An accelerator driven electron bunch passes through an undulator, co-propagating with a light pulse trapped in an optical cavity. The light pulse is amplified in the undulator and reflected back and forth by a set (typically a pair) of cavity mirrors so that at the entrance of undulator it meets the electron bunch (either a new bunch from a linac or the same bunch circulating in a storage ring \cite{30yj}) again. With $g$ is the single-pass gain (relative increase of the optical intensity per pass) and $r$ is the cavity round-trip reflectivity related to the active output coupling and passive cavity loss, if $(1+g)r>1$, the light pulse would evolve from the initial incoherent spontaneous emission to a coherent pulse. After an exponential growth, the gain decreases at high intra-cavity optical intensity due to over modulation, and when $(1+g)r=1$, the system reaches a steady state, i.e., saturation.

\begin{figure}
  \centering
  \includegraphics[width=8cm]{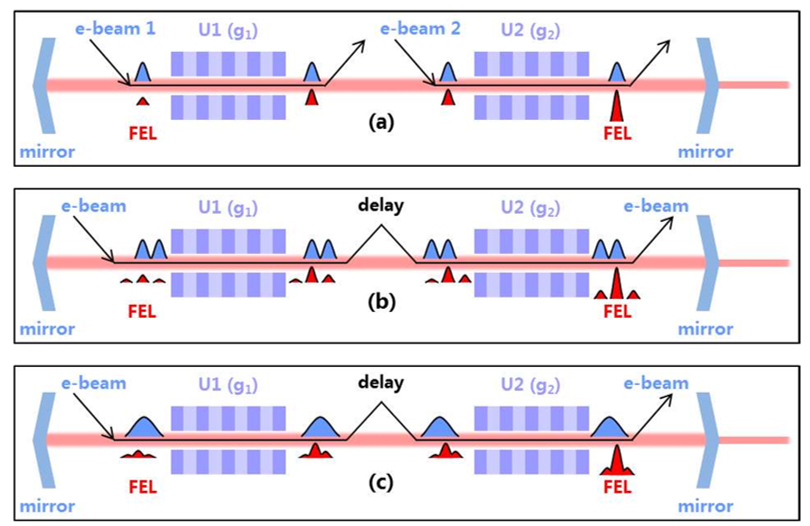}\\
  \caption{(color online). Schemes for gain cascading in single-pass of FEL oscillators. (a) The original idea conceptually to replace the degenerated electron bunch with a fresh one. (b) A more practical method of using bunch trains and refresh it through the delay between the two stages of undulators. (c) The bunch trains are replaced by a single long electron beam.}\label{scheme}
\end{figure}

In general, the electron-to-light energy conversion efficiency can be evaluated as following: in order to get highest output power, which means that most electrons move to the bottom of the bucket in the phase space and contribute maximum energy, the number of undulator periods and the reflectivity of cavity mirrors are carefully optimized. At saturation, the electron pulse will generate radiation energy equals to the amount of energy coupled out of the cavity in a single pass. Assuming that electron beam on the average lose energy equals to half of the bucket amplitude which can be calculated by pendulum equation \cite{30hz}, so that the generated radiation power is $P_{out}=P_{beam}/4N_u$. Meanwhile, to start the exponential growth more easily, a larger single-pass gain $g$ (larger $N_u$) is preferred. Thus, the design optimization of an FEL oscillator is a trade-off between increasing single-pass gain and enhancing lasing peak power \cite{30qb}. To enhance the output peak power of an FEL oscillator, as illustrated in Fig.~\ref{scheme}(a), it is straightforward conceptually to use multiple electron beams in one round trip. For such an oscillator which employs low gain theory with $n$ undulator stages, or $n$ electron beams, we ignore the second order small quantity and the single-pass gain will be roughly the sum of these undulator stages, and the FEL output power should be written as $P_{out}=\sum\limits_i^n \left( P_{beam}/4N_i \right)$, where $N_i$ is the period number of the $i$-th stage undulator. For a given required single-pass gain for FEL oscillator, the more undulator stages are employed, the smaller $N_i$ is in each stage, which leads to an extra enhancement of output peak power, thus cascading FEL oscillator could generate pulses with peak power increases by a factor larger than $n$.

\begin{figure}
  \centering
  \includegraphics[width=8cm]{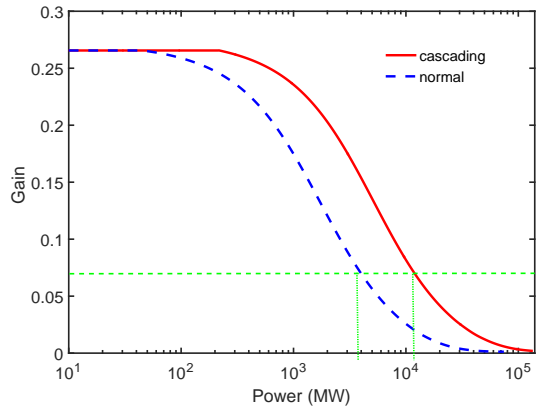}\\
  \caption{Single-pass gain as a function of cavity pulse power. The dashed green line represents the single-pass gain at saturation for the typical 6\% output coupling. For the same output coupling efficiency, the gain cascading FEL oscillator generates a larger light pulse power inside the cavity than the traditional FEL oscillator.
  }\label{gain}
\end{figure}

The single-pass gain of this new scheme is calculated using the theory in \cite{30lk}. Fig.~\ref{gain} shows the gain as a function of different laser peak power inside the cavity. The gain remains constant at low light power, and it drops off gradually as the light pulse energy increases. The gain reducing and FEL oscillator saturation is due to the over modulation of electron beam in the undulator. In the traditional FEL oscillator, the electron beam interacts with pulse in one section of relatively longer undulator and the energy spread increases continually. Thus even if the initial single-pass gain is same at the beginning, it decreases earlier and faster as shown in Fig.~\ref{gain} at the saturation regime. In the gain cascading FEL oscillator, however, the degenerate electron beam is replaced with a new bunch or delayed to a fresh part between the two undulators, and then interact with laser pulse further more. It gives out the same single-pass gain at first, decreases at a higher pulse power and drops off more slowly in the saturation region due to the relatively shorter length of undulator in each stage. In other words, the cascading FEL oscillator takes the advantages of slower decrease of gain curve for the shorter undulators, and uses multi-stages of them to get enough single-pass gain to ensure the growth of initial spontaneous radiation. As a consequence, the light pulse obtains more energy from the beam at each round trip in the saturation and FEL oscillator generates higher power as demonstrated in Fig.~\ref{gain}.

The challenges of an FEL oscillator shown in Fig.~\ref{scheme}(a) include: first, a complex accelerator which is able to generate multiple electron beams and switch them to the specified undulator stage; and second, a tight arrival timing jitter of these electron beams for accurate synchronization. Therefore, two other schemes relying on experimentally demonstrated accelerator techniques and elements are proposed in Fig.~\ref{scheme}. Scheme (b) operates with twin-bunch and/or multi-bunch, which can be generated with different methods: inserting a designed slotted foil between the second and third dipole of the bunch compressors of the accelerator \cite{18ep}, generating multi-pulse beam by shaping the temporal distribution of photocathode laser with alpha barium borate ($\alpha$-BBO) crystals \cite{31mp,32ff} or auto-correlating two chirped laser pulses \cite{33sy}, and taking advantages of longitudinal space charge forces to magnify the initial density modulation can produce high peak current bunch trains \cite{34zz}. The bunch length and time interval between bunches can be flexibly tuned, and then using several appropriate chicane delays, it is easy to make the central light pulse meet a new bunch in the next stage. Then after all stages, the single-pass gain of the central light pulse is maximum compared with that of the other light pulses. Under such circumstances, when the central light pulse saturates, the intensity of other pulses is still negligible. Considering the sequence of the light pulses that each electron bunch interacts with, after saturation of the central pulse, those light pulses behind the central one gain little because of the beam quality dilution, while the others still gain normally. Therefore, to prevent the buildup of multiple light pulses and ensure the desired growth of the central pulse in the cavity, a criterion is derived as following,
\begin{equation}
r \prod_1^n (1+g_i)> 1
\label{eq1}
\end{equation}
\begin{equation}
r \prod_1^{n-1} (1+g_i)< 1
\label{eq2}
\end{equation}
The first formula in Eq.~(\ref{eq1}) ensures that the part of radiation overlaps with electron bunch at every stages of undulators gets enough gain and the power grows up. In the Eq.~(\ref{eq2}), the left term is the net single-pass gain of the light pulse next to the central one, which is not sufficient to grow up in the oscillator. The other parts of radiation get even less gain and does not increase at all.

 Scheme (c) comes up from the concept of cascading seeded FEL with fresh-bunch technique \cite{34wj}. In the first undulator section the whole electron beam generates spontaneous emission as a seed, however, with an appropriate choice of system delay parameters, only one specified part of the light pulse overlaps with electron beam and gains in all undulator stages and saturates finally. According to Eq.~(\ref{eq2}), the other parts of radiation cannot get enough single-pass gain and does not grow. In this way, the output laser pulse is equivalent to that of a shorter electron bunch, however the peak power is nearly the sum of several of them. To make the fully use of the whole electron beam, we choose the summation of all the delay length plus the length of radiation pulse is equal to the electron bunch length. For the special case when all the $n-1$ delay length is identity the same, which means the length of light pulse generated is $1/n$ of the classical case (with $N_u$ undulator periods), one can calculate the extraction efficiency ratio
 \begin{equation}
\frac{\varepsilon_{cas}}{\varepsilon_{norm}} = \frac{\left(\sum\limits_i^n \left( 1/4N_i \right) \right) /n}{1/4N_u}
\label{eq3}
\end{equation}

It is well suited for the cases that the slippage length is much less than the electron bunch length, i.e., X-ray FEL oscillators \cite{35kkj,36lrr,37dj}. It is worth stressing the main difference between our proposals and the optical klystrons here. The chicanes in our schemes are mainly used to control the timing between the electron bunch and the light pulse. The typical delays here are at the time scale of picoseconds, which are large enough to smear out the electron bunching created at previous undulator stages, and ensures the small-signal gain in all stages. In comparison, the chicane in optical klystron \cite{38vna,39dib} mainly provides a longitudinal dispersion which transfers the energy modulation in previous undulator stage to density bunching, and thus offers a high single-pass gain for an oscillator.

\section{Simulation of infrared FEL oscillator}
\begin{figure}
  \centering
  \includegraphics[width=8cm]{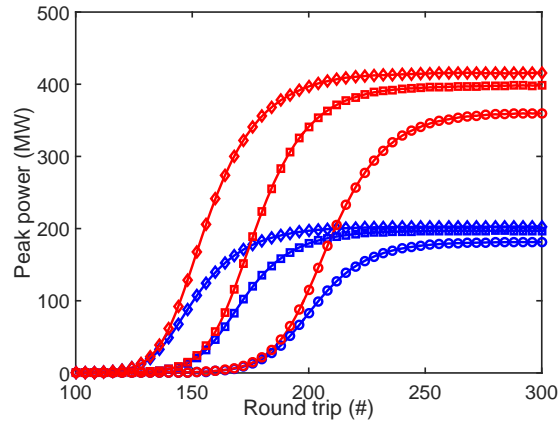}\\
  \caption{ (color online). The growths of output peak power in FEL oscillators from steady-state simulations, in which the blues are results from a classical oscillator with 25\% single-pass gain from a single-stage undulator, while the reds are results from an FEL oscillator with 25\% single-pass gain from two-stage cascaded undulator, i.e., ~12\% from each one. The diamond, square and circle represent the cases with the output coupling efficiency of 6\%, 8\% and 10\%, respectively, and the passive loss of the cavity is assumed to be 1\%.}\label{IR_round_trip}
\end{figure}

To show the benefits of the gain cascading in the proposed schemes, an example with two cascading undulator stages of scheme (b), with 12\% gain in each stage, is illustrated for lasing at 1.6 $\mu$m. The propagation in free space, reflection and focusing of light pulses are modeled using OPC \cite{40pjm}, and FEL gain in undulator stages are simulated with the code GENESIS \cite{41rs}. For simplicity, the parameters of each electron bunch and undulator stage are assumed to be the same in this paper: the beam energy of 80 MeV, normalized transverse emittance of 10 $\mu$m-rad, relative energy spread of 0.2\%, electron bunch peak current of about 200 A with full width at half maximum (FWHM) bunch length of 0.5 ps and bunch charge of 100 pC, time interval between two bunches of 1.0 ps; the undulator period length of 45 mm, period number of each undulator stage of 14. Regarding the parameters of FEL cavity used, the Rayleigh range and cavity length are 0.35 m and 6 m, respectively. All the parameters can be routinely realized for modern accelerators \cite{41ngr}. Fig.~\ref{IR_round_trip} plots the peak power as a function of the pass number in a steady-state simulation. The FEL oscillator saturates with a stable output peak power around 400 MW after 200-250 passes, depending on the output coupling efficiency of the cavity (the ratio of output power to intra-cavity power). In comparison, the growth curves of a classical FEL oscillator with 25\% single-pass gain are presented (with 16 undulator periods), which results a 200 MW peak power. Considering the roughness of the theoretical predictions, it is reasonably consistent with the steady-state simulation results.

\begin{figure*}
  \centering
  \subfigure{\includegraphics[width=6.2cm]{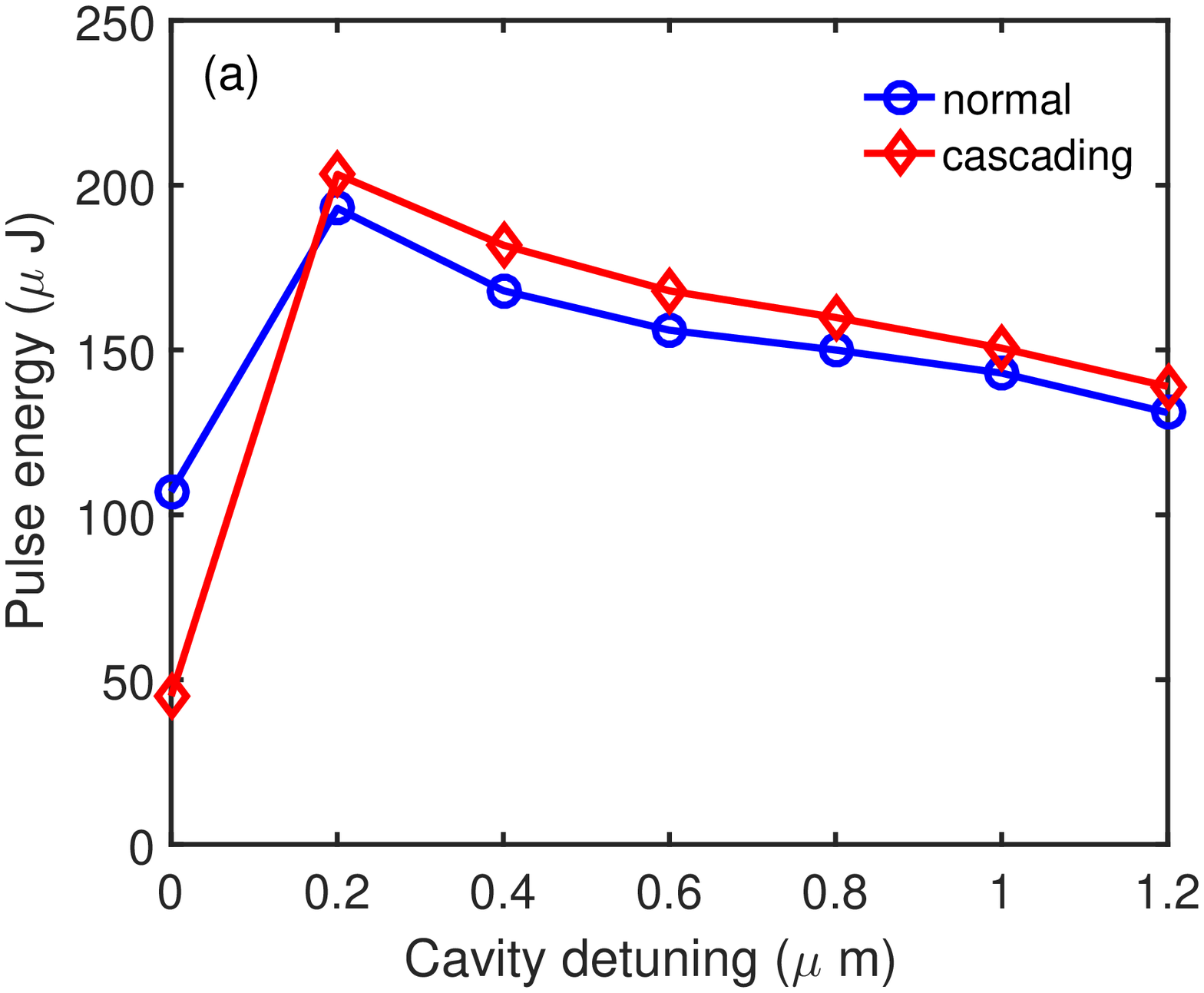}}
  \subfigure{\includegraphics[width=6.3cm]{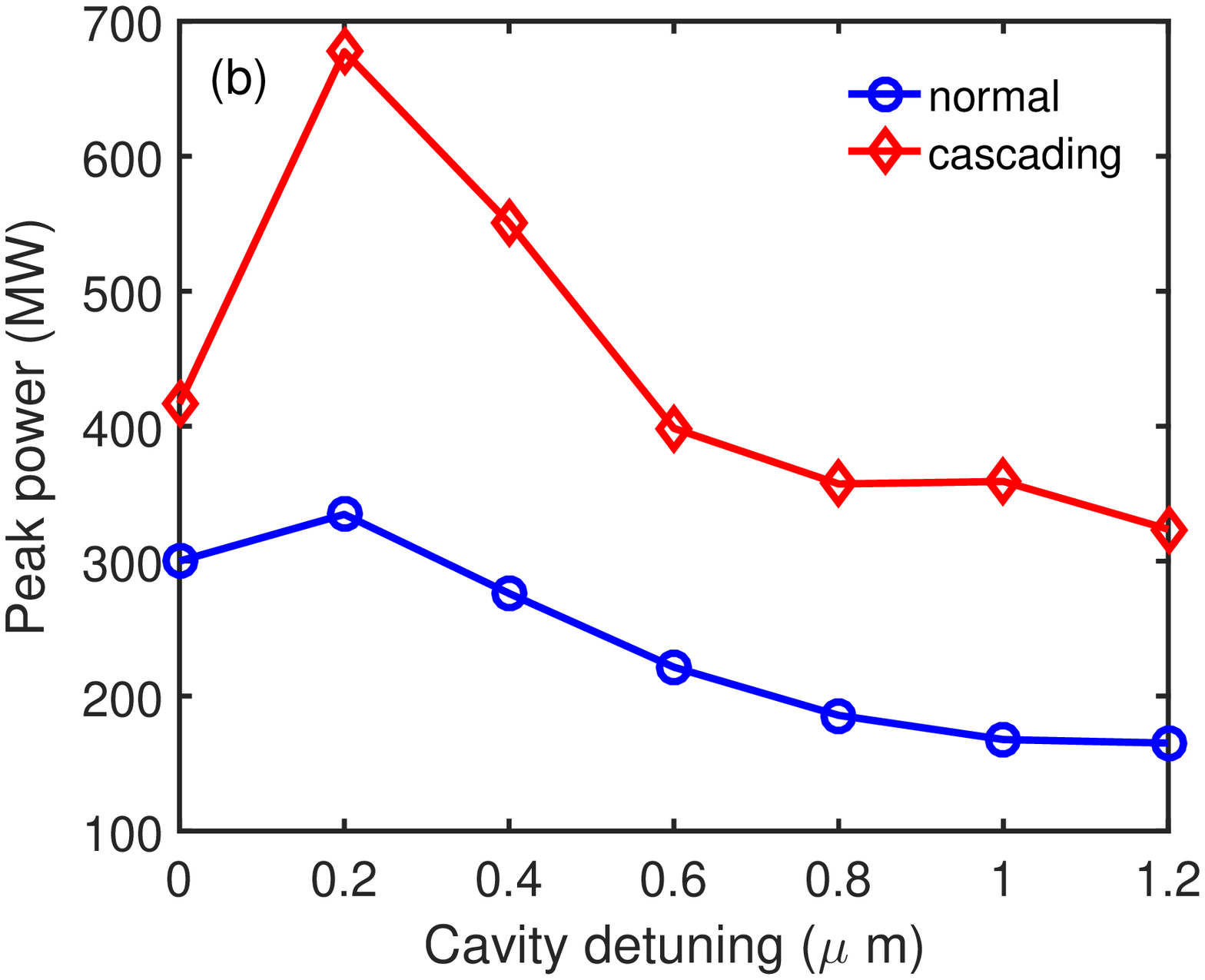}}
 \caption{\label{detune} (color online). The cavity detuning curve for (a) output energy and (b) peak power.}
\end{figure*}
\begin{figure*}[t]
  \centering
  \subfigure{\includegraphics[width=3.8cm]{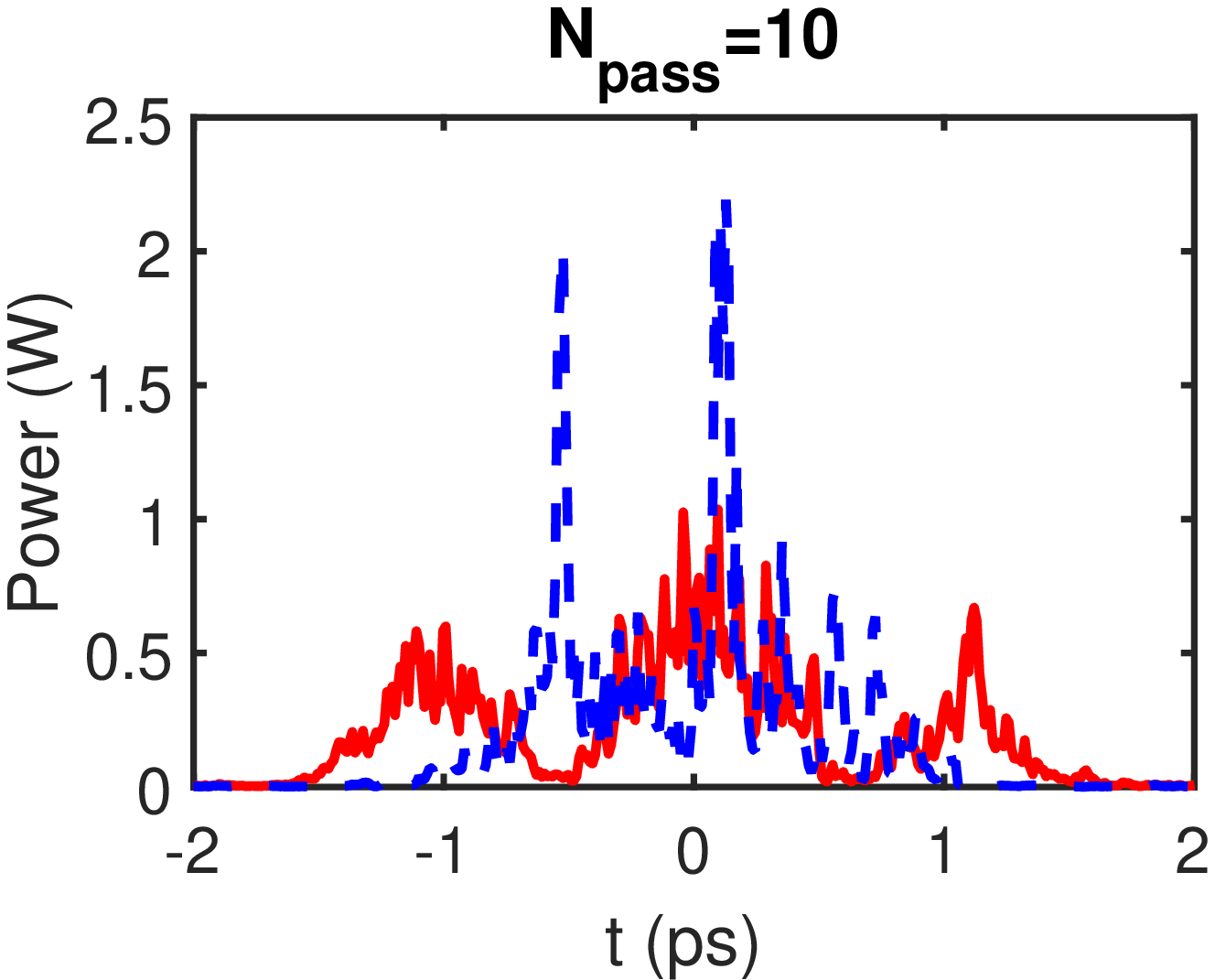}}
  \subfigure{\includegraphics[width=3.8cm]{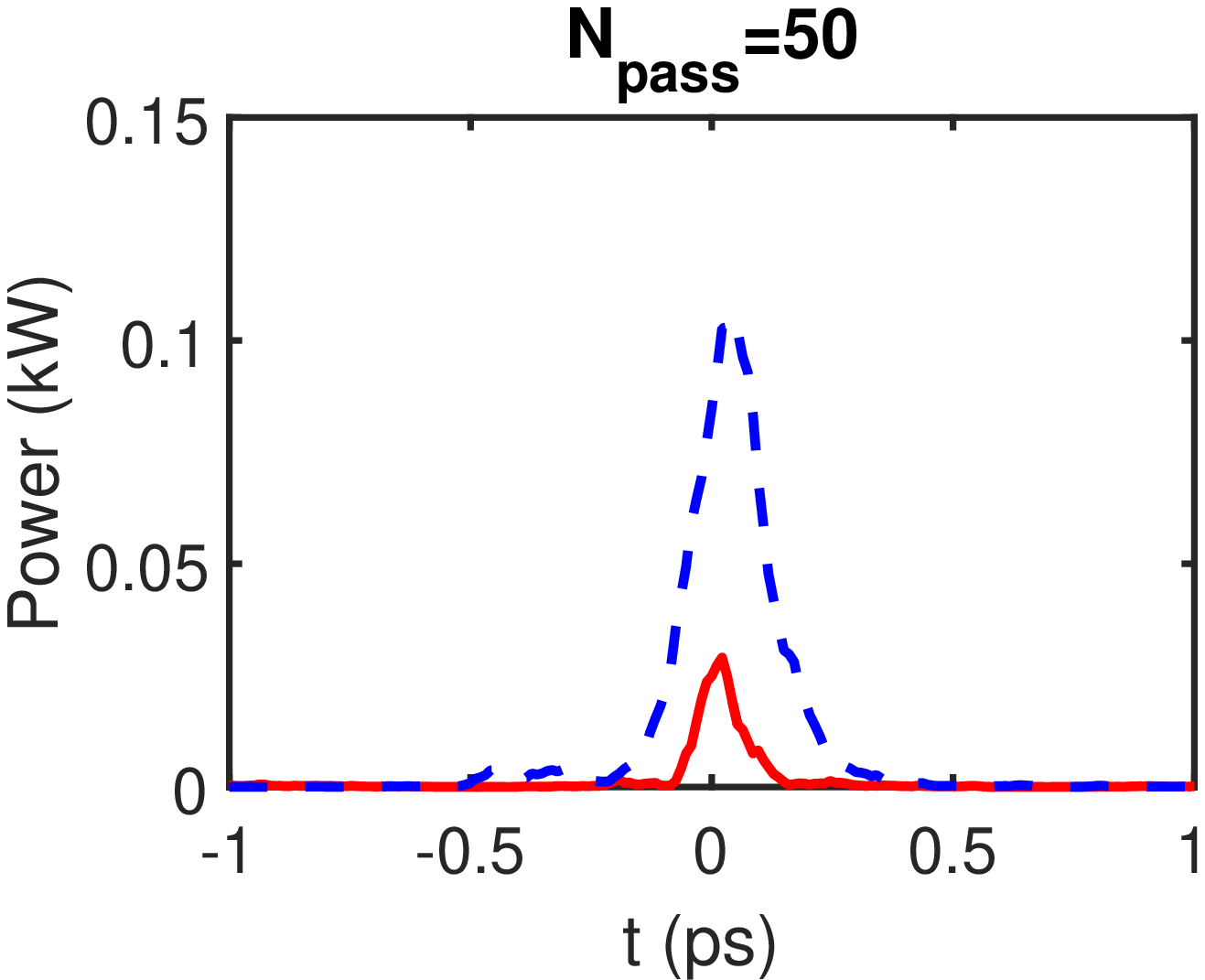}}
  \subfigure{\includegraphics[width=3.8cm]{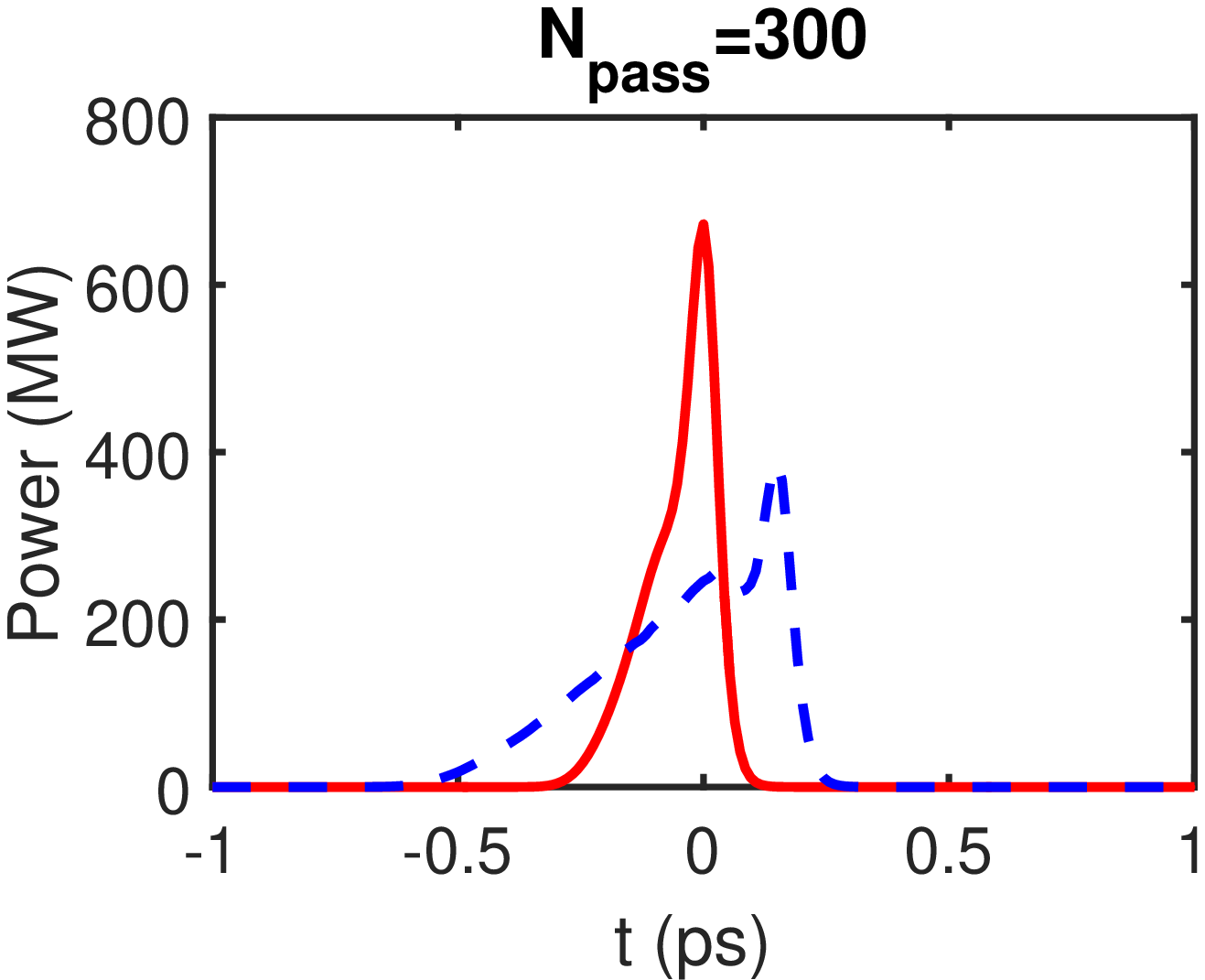}}
  \subfigure{\includegraphics[width=3.8cm]{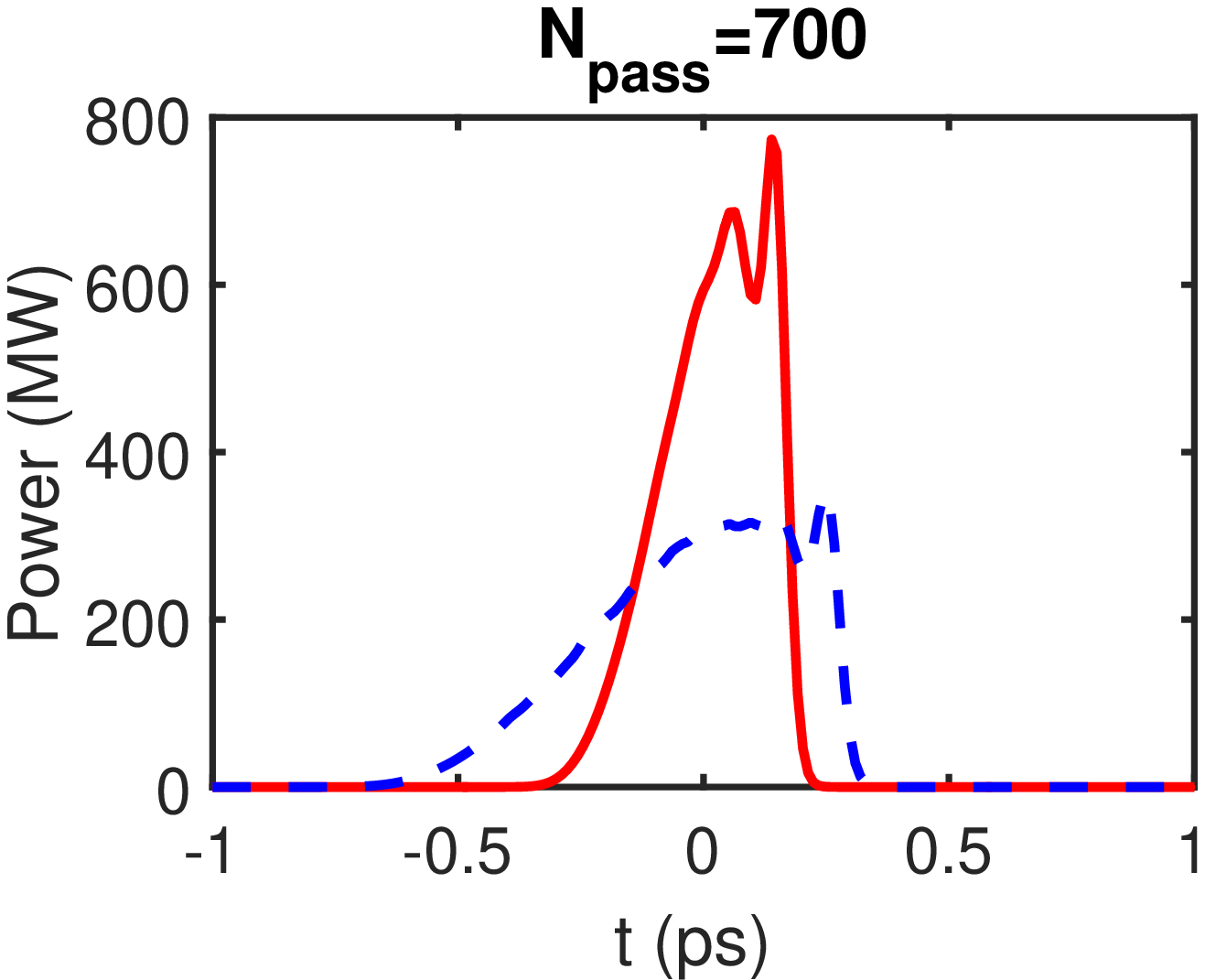}}
  \subfigure{\includegraphics[width=3.8cm]{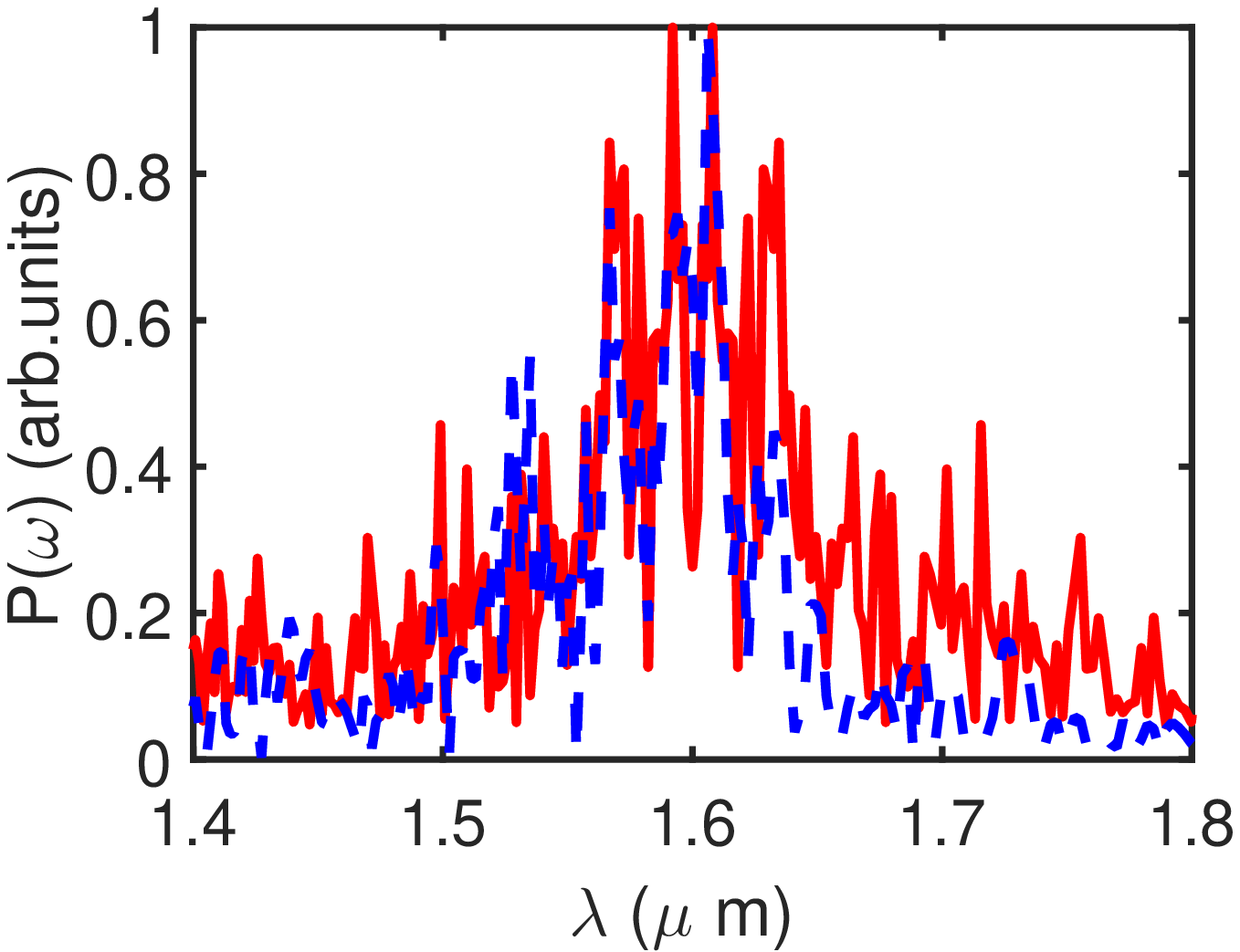}}
  \subfigure{\includegraphics[width=3.8cm]{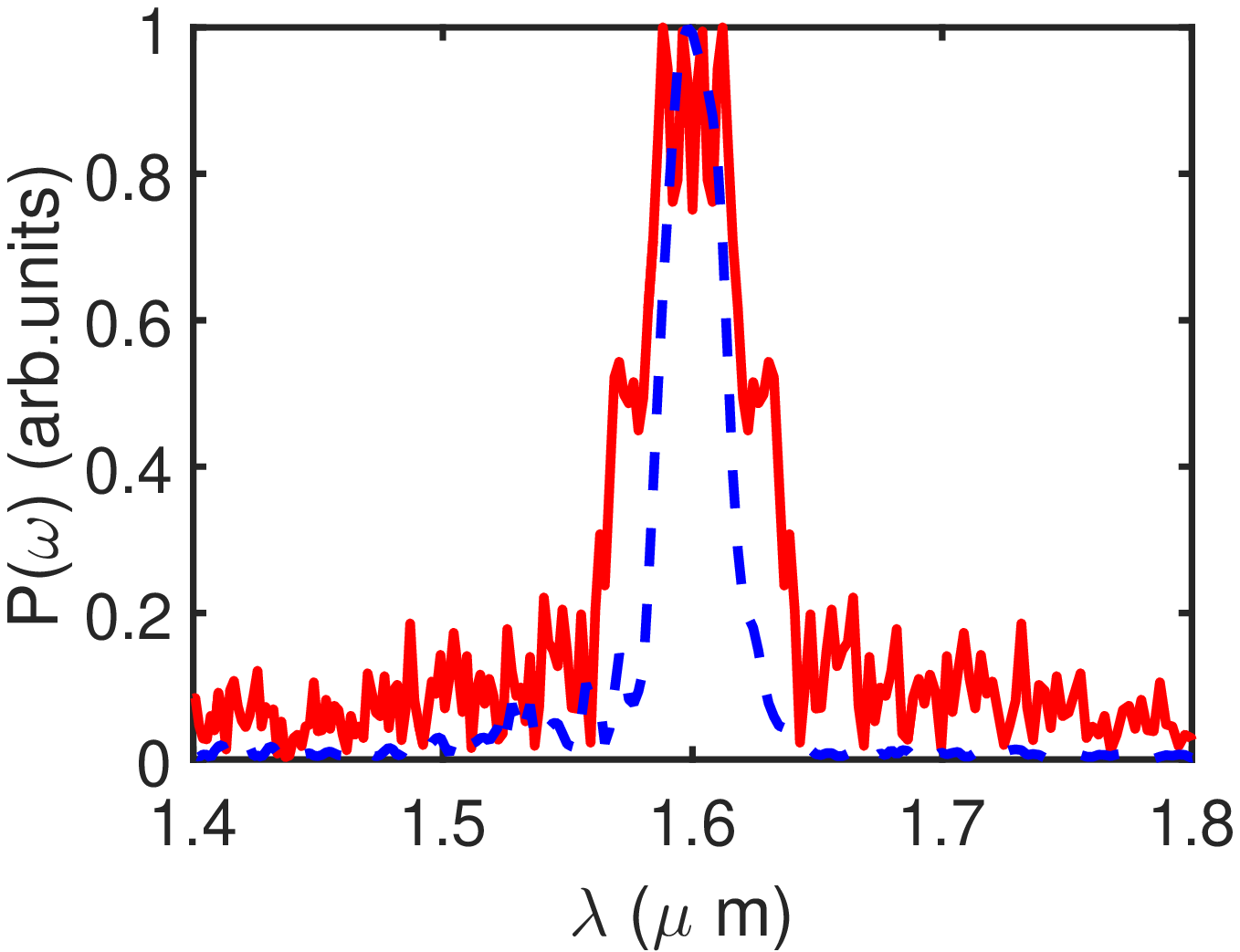}}
  \subfigure{\includegraphics[width=3.8cm]{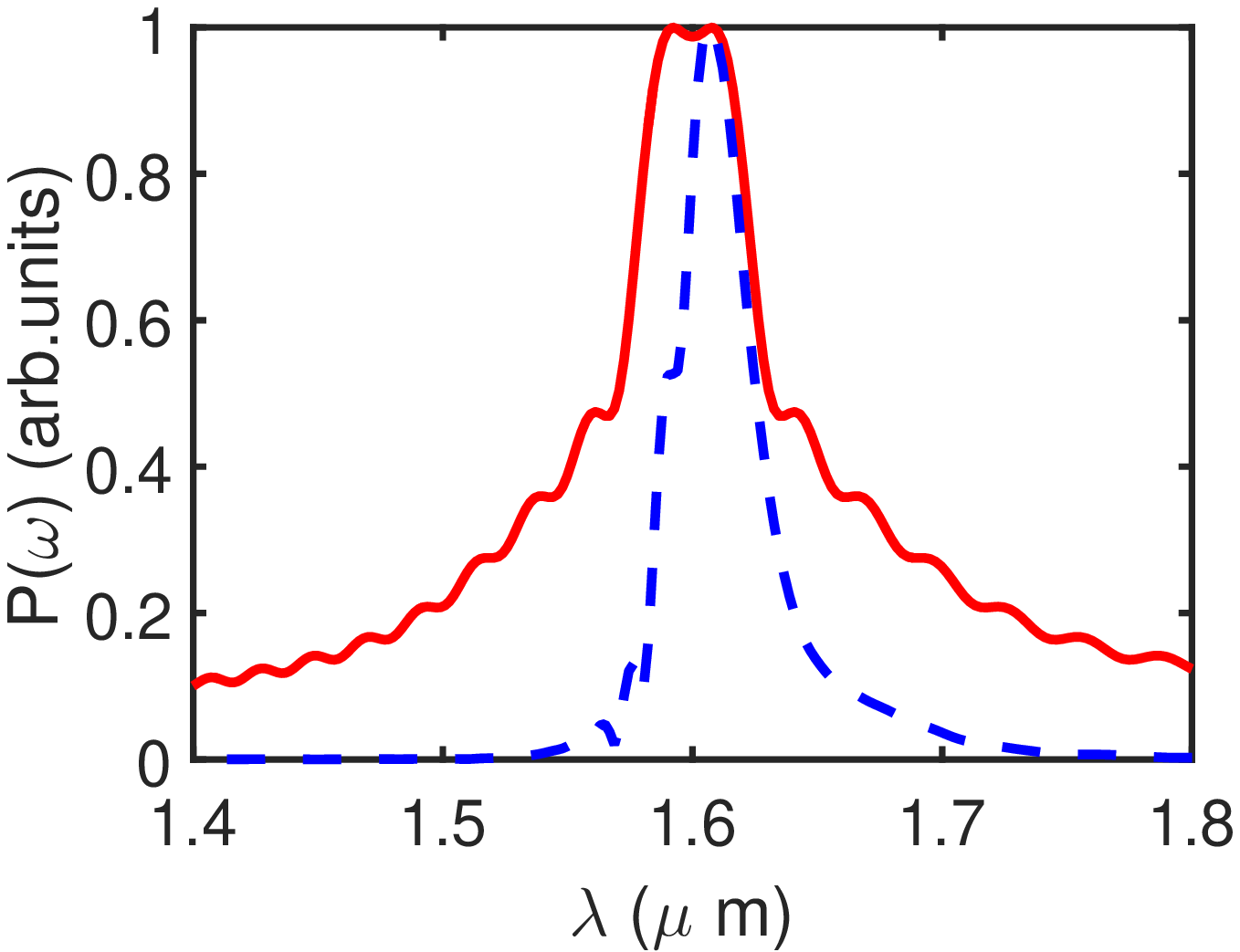}}
  \subfigure{\includegraphics[width=3.8cm]{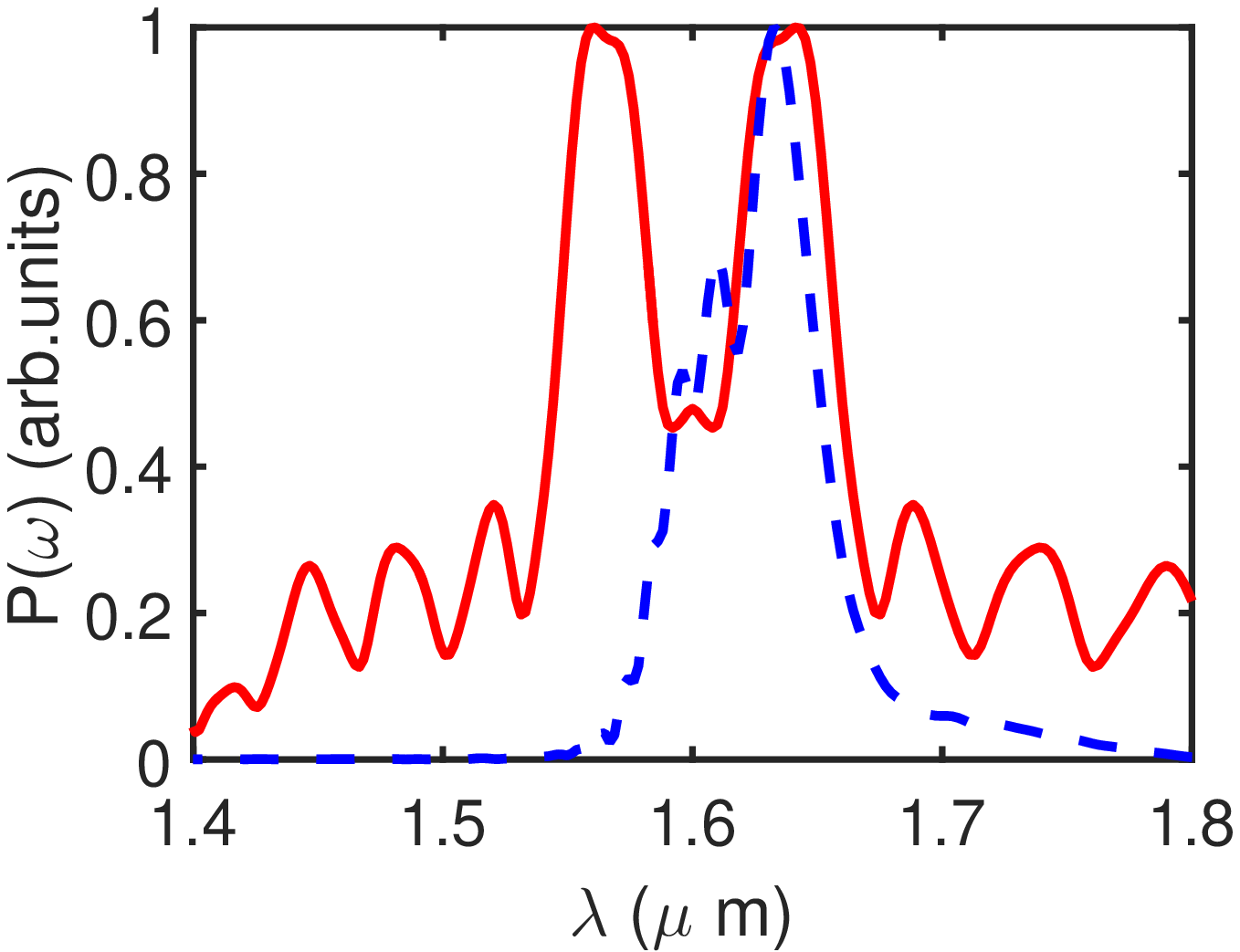}}
 \caption{\label{IR_profile} (color online). The temporal and spectral profile of the output 1.6 $\mu$m pulse in the conventional FEL oscillator (blue dashed line) and the two-stage undulator cascaded FEL oscillator (red solid line), with the chicane delay of 1.0 ps. The output coupling efficiency is 6\% and the cavity detuning length of 0.2 $\mu$m for both of them. (The upper figures is manually shifted to the center.)}
\end{figure*}

With the optimized parameters from steady-state simulation, time-dependent simulations are carried out for taking into account the cavity detuning effects \cite{42JD} and the FEL temporal structures. The 1.6 $\mu$m FEL oscillator output power with different cavity detuning is shown in Fig.~\ref{detune}. It demonstrates that the cascaded FEL oscillator produces twice the peak power as well as nearly the same pulse energy with classical FEL oscillator, which agrees well with the steady-state simulation and Eq.~(\ref{eq3}). The maximum output energy is achieved at 0.2 $\mu$m cavity detuning, and Fig.~\ref{IR_profile} shows the output pulse performance. The first row of panels displays the evolution of temporal power profile with the increasing pass number, while the second row shows the corresponding spectra. According to the simulation, 1.6 $\mu$m radiation with 677 MW peak power can be generated in the oscillator with two-stage cascaded undulators. The total energy per pulse coupled out of the cavity is 203 $\mu$J. Fig.~\ref{IR_profile} demonstrates a temporal width of 352 fs (FWHM) and a spectral width of 114 nm (FWHM) for the output FEL pulse. For comparison, the normal FEL oscillator is simulated using the same peak current electron beam with twice the pulse duration, which provides the same charges as the cascaded FEL oscillator. It gives out the maximum 193 $\mu$J laser pulse energy with 337 MW peak power when cavity detuning equals to 0.2 $\mu$m. The results consistent with Eq.~(\ref{eq3}), which reveals that pulse energy should be approximately the same and cascaded FEL oscillator output peak power is nearly twice as much as that of classical FEL oscillator.

\section{Simulation of X-ray FEL oscillator}
To examine the enhancement of peak power, Scheme (c) is also simulated using an XFELO lasing at 1 $\rm{\AA}$. Since the electron bunch length is far larger than the lasing wavelength, the delay due to slippage effect is ignored, and an XFELO with four-stage cascaded undulators is used. According to Ref.~\cite{35kkj}, we choose parameters of XFELO: 7 GeV beam energy, 1.4 MeV energy spread, 0.2 $\mu$m-rad normalized emittance, 18 m average $\beta$ function, and 20 pC bunch charge with flat-top current of 10 A, are used in the simulation. A quasi-confocal X-ray cavity of 100 m cavity length and cavity crystal mirror with 10 meV Bragg reflection bandwidth are employed. The total reflectivity, output coupling and passive loss of the XFELO cavity are assumed to be 89.8\%, 4.5\% and 1\%, respectively. According to Eq.~(\ref{eq1}) and Eq.~(\ref{eq2}), the gain at each undulator stage fulfills $3\% < g_i < 4\%$, so that undulator parameters 18.8 mm$\times$960 per stage are chosen, which provide gains of 4.0\%, 3.5\%, 3.5\% and 4.0\% respectively. For comparison, a classical no-delay XFELO with the same parameters is simulated, the undulator periods is 2000 in order to maintain the same nearly 15\% single-pass gain with cascading XFELO.

\begin{figure*}[t]
  \centering
  \subfigure{\includegraphics[width=3.8cm]{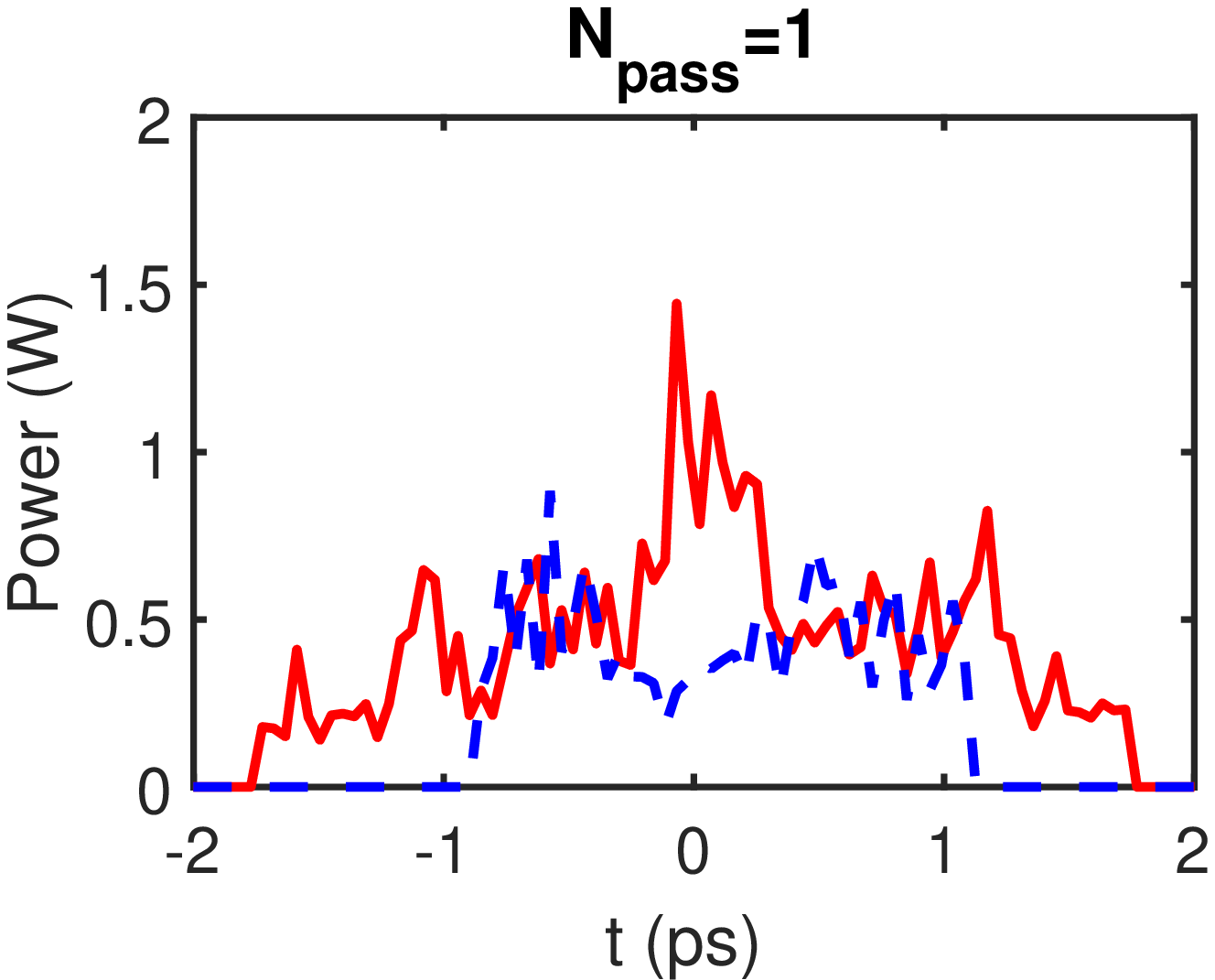}}
  \subfigure{\includegraphics[width=3.8cm]{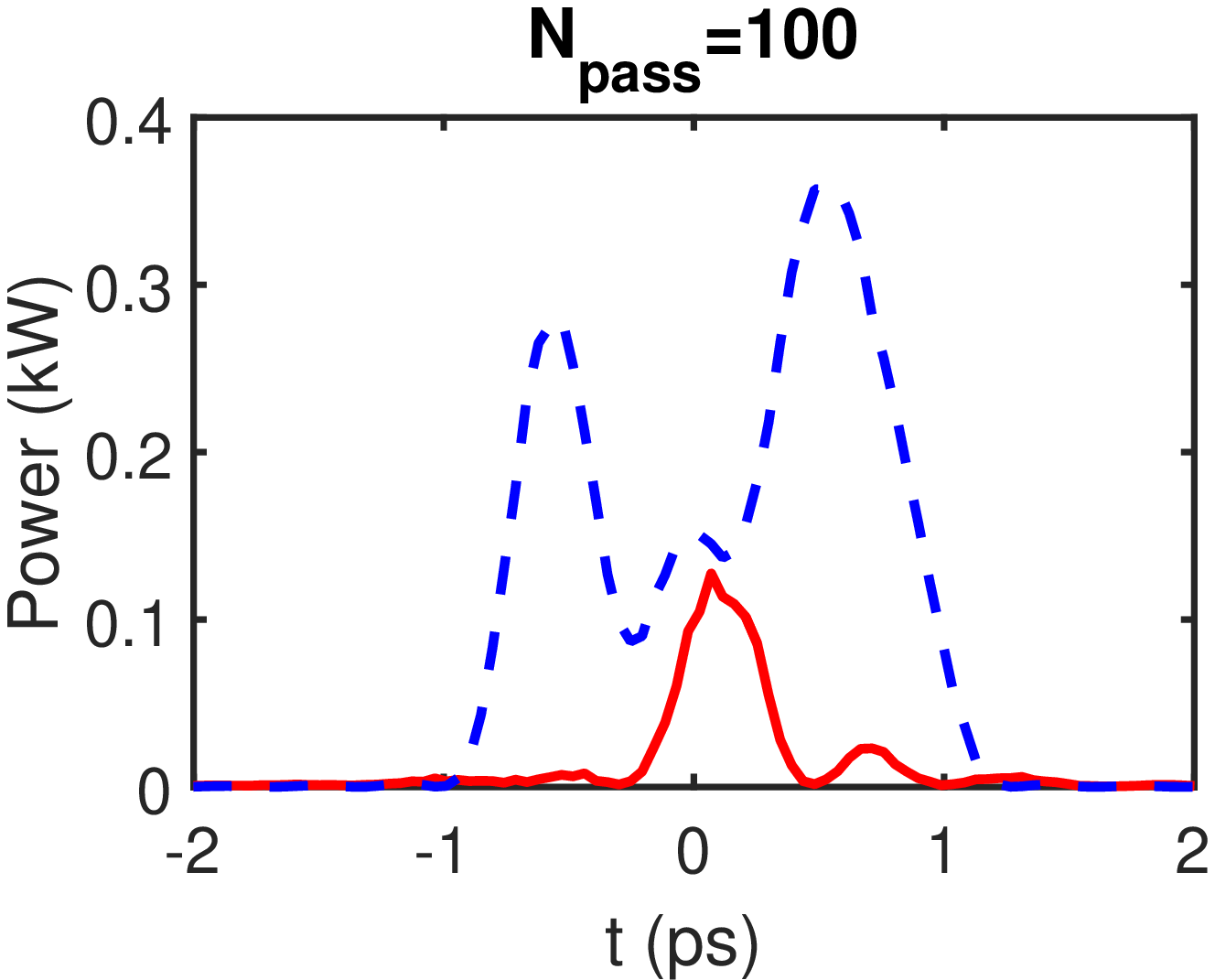}}
  \subfigure{\includegraphics[width=3.8cm]{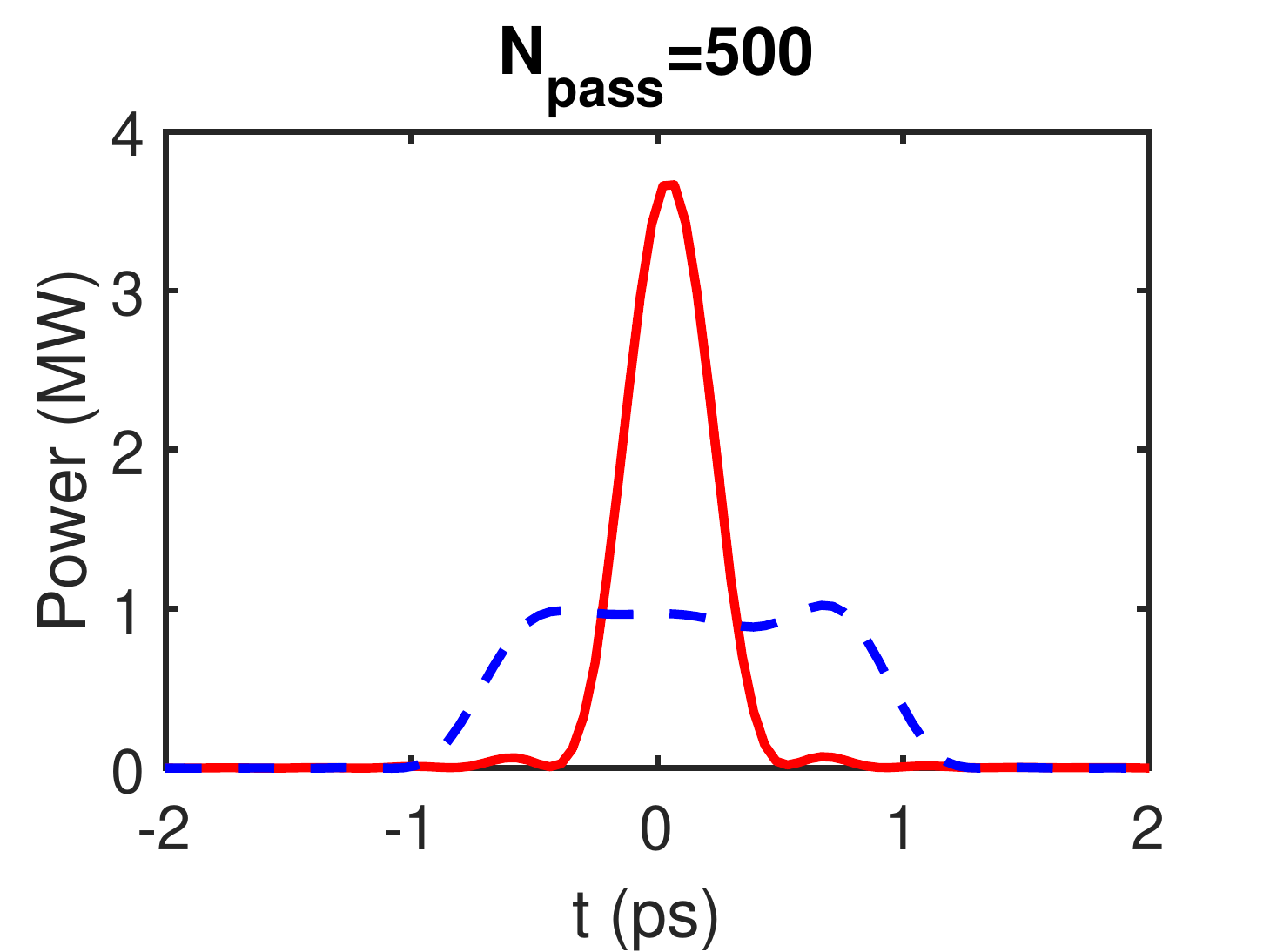}}
  \subfigure{\includegraphics[width=3.8cm]{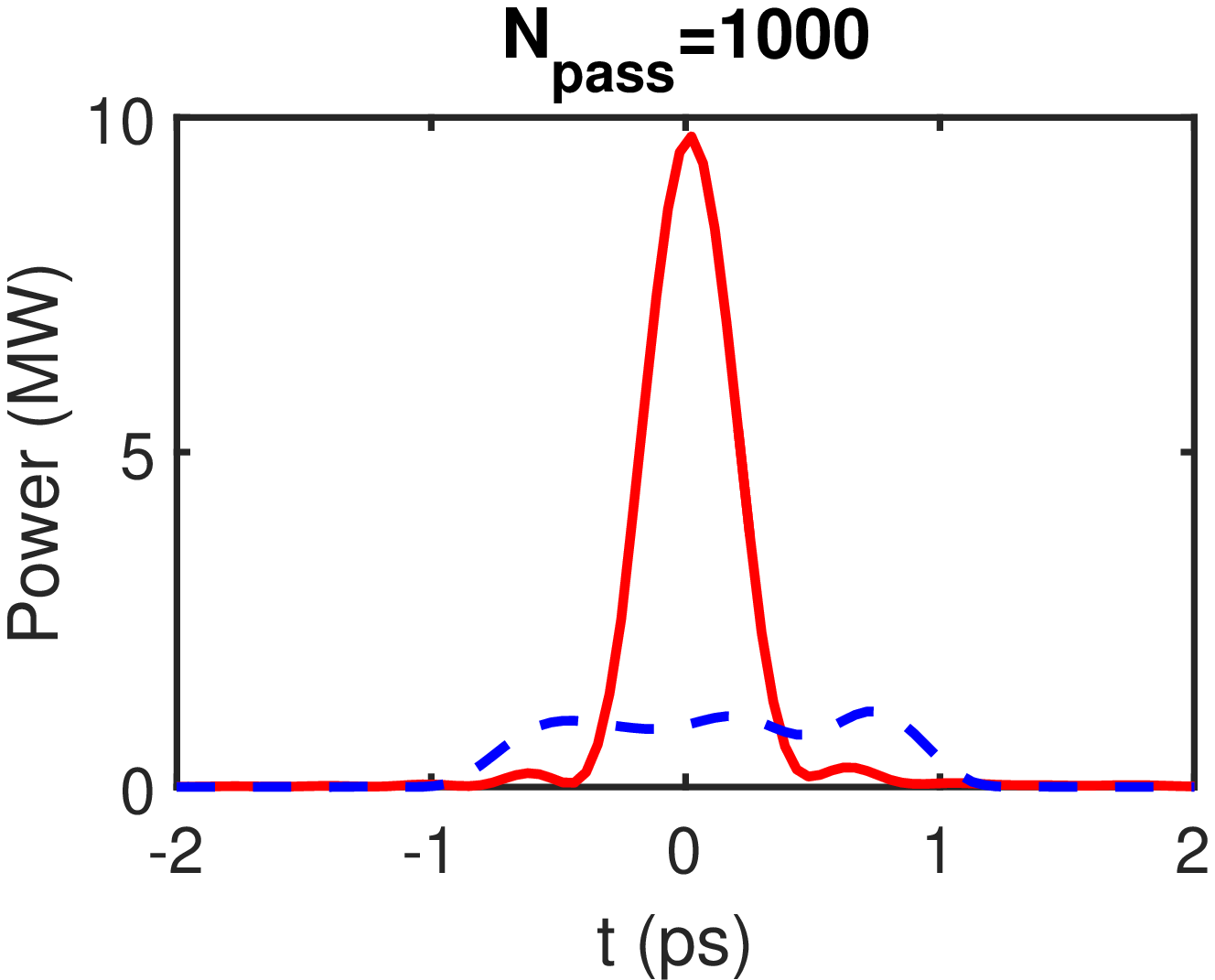}}
  \subfigure{\includegraphics[width=3.8cm]{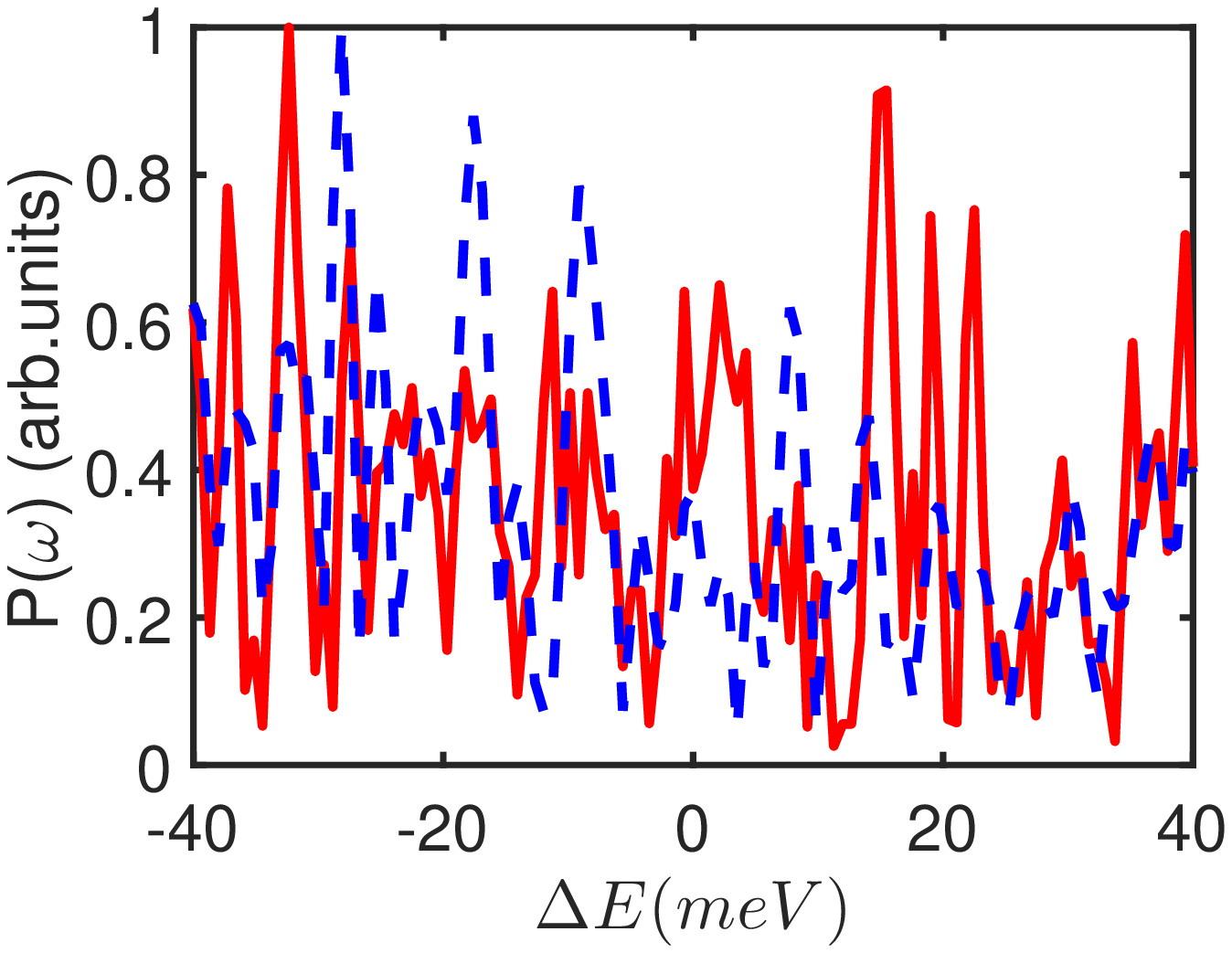}}
  \subfigure{\includegraphics[width=3.8cm]{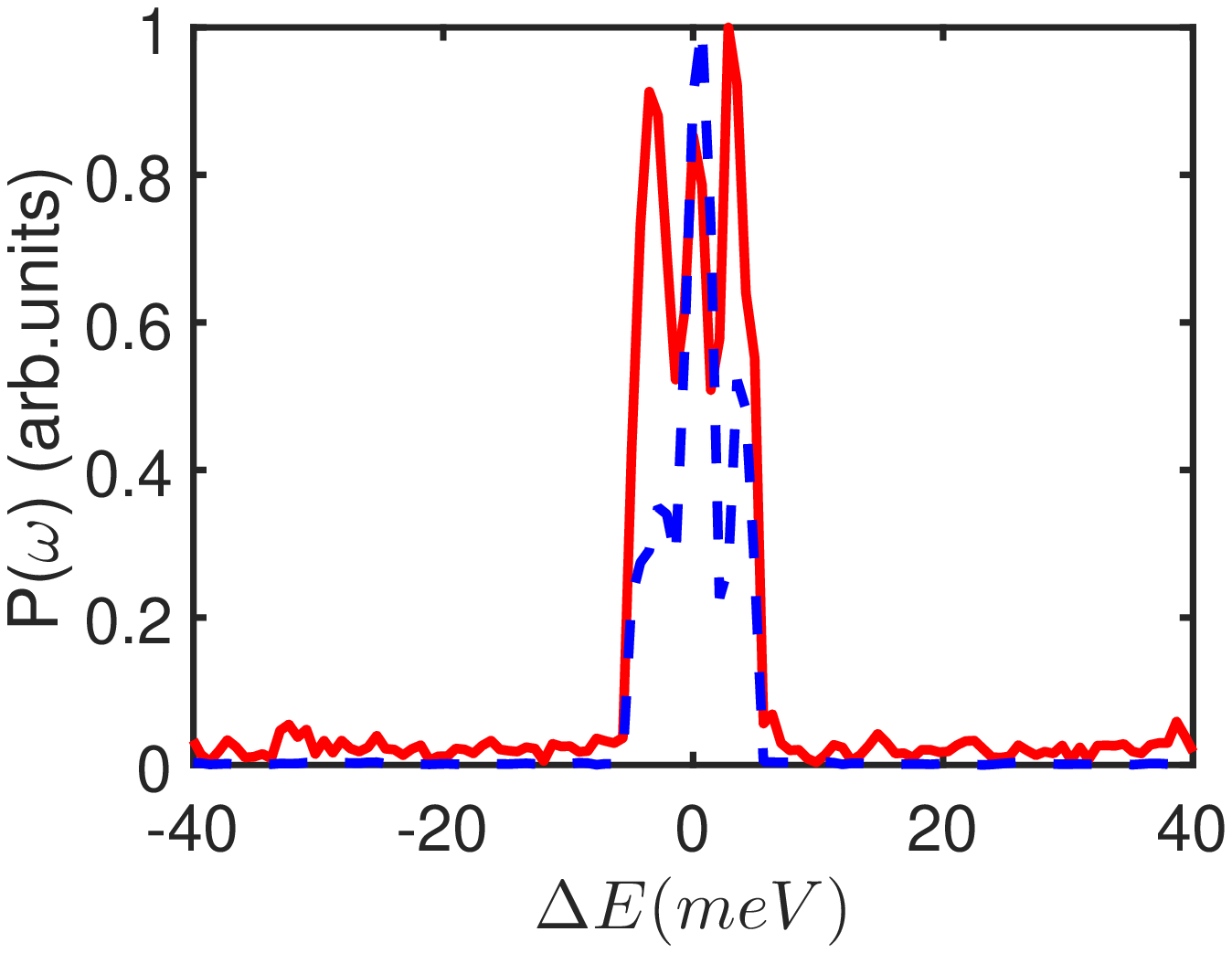}}
  \subfigure{\includegraphics[width=3.8cm]{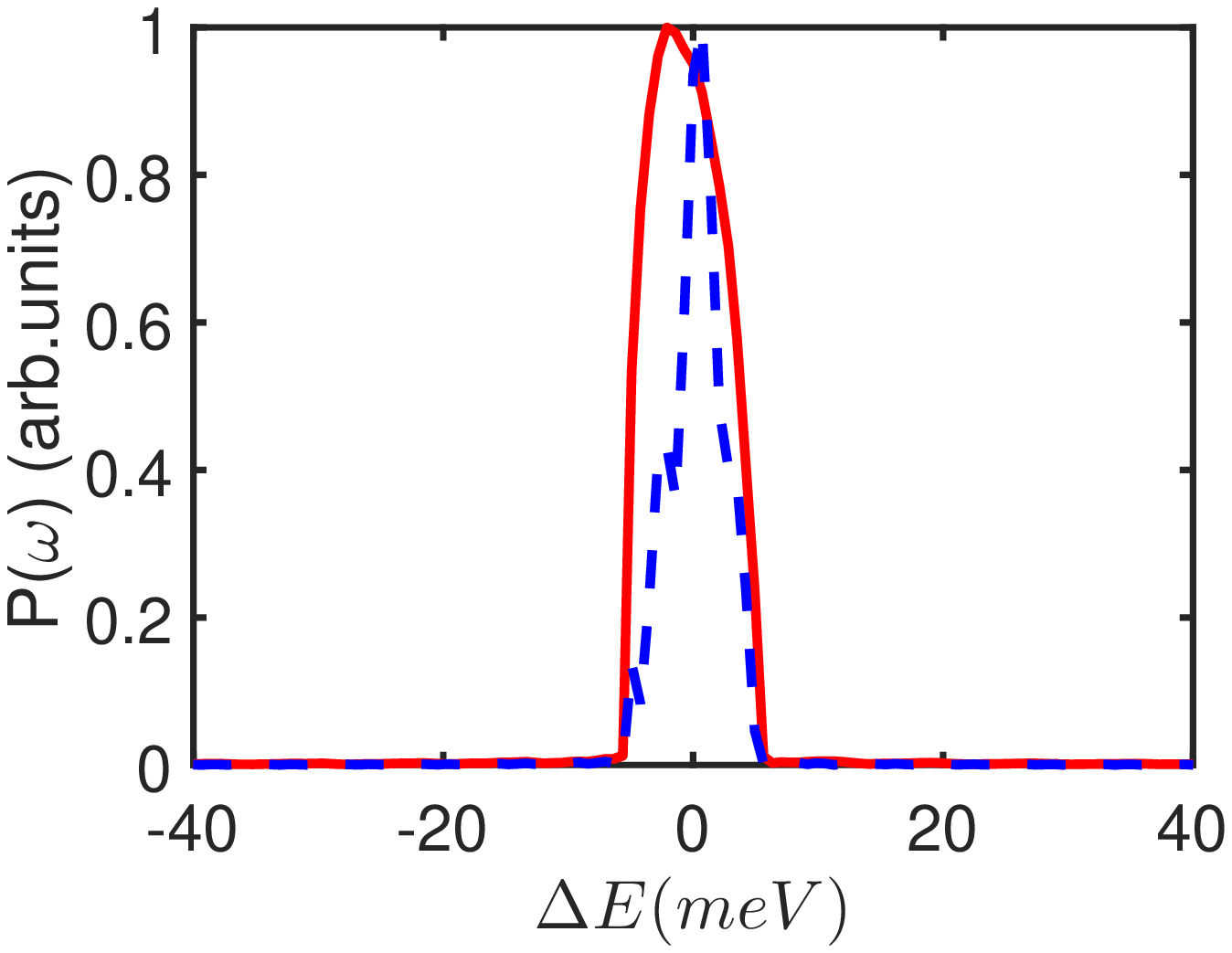}}
  \subfigure{\includegraphics[width=3.8cm]{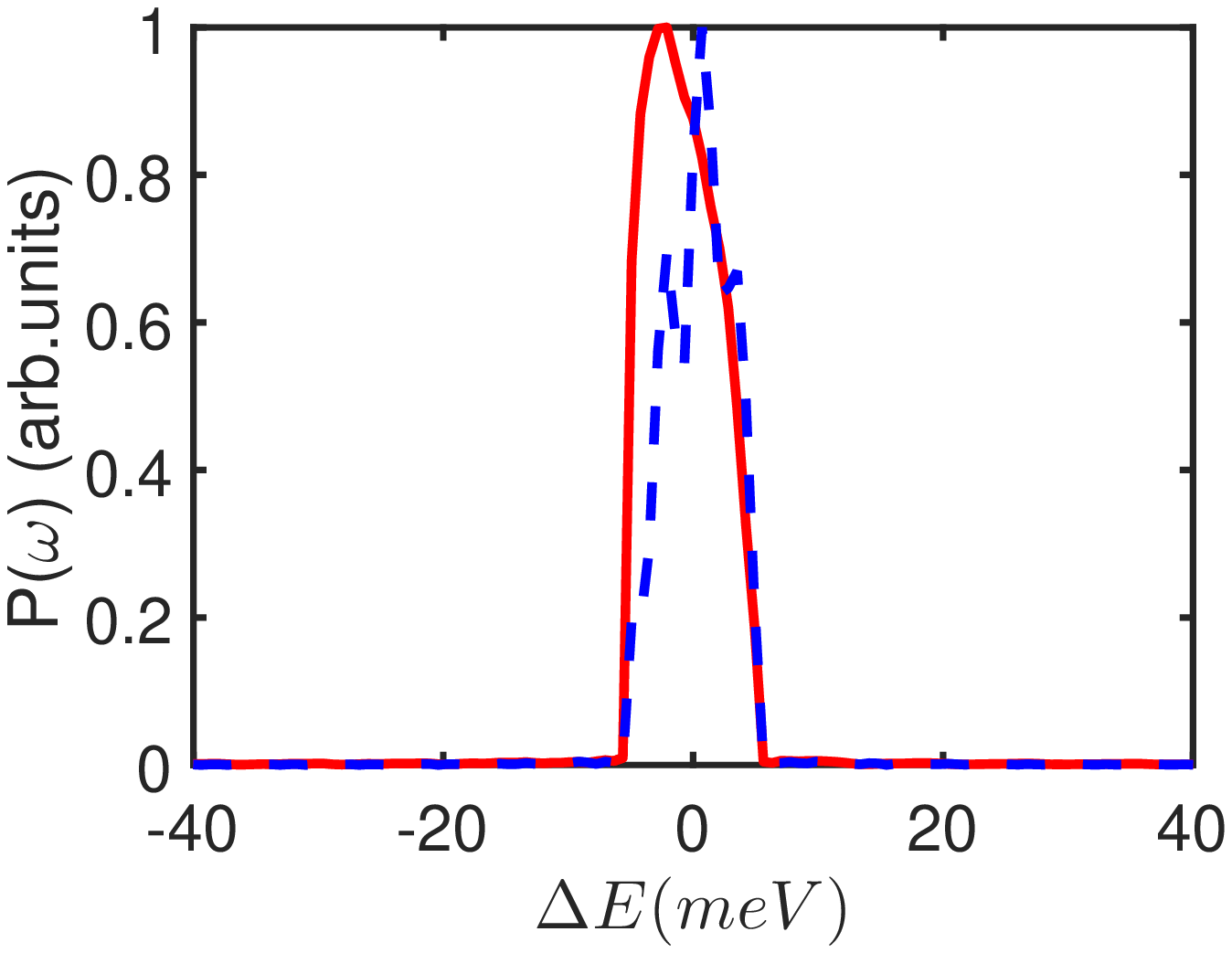}}
 \caption{\label{x_profile} (color online). The temporal and spectral profile of the output $1 \rm{\AA}$ pulse in the normal FEL oscillator (blue dashed line) and four-stage undulator cascaded FEL oscillator (red solid line).}
\end{figure*}

\begin{figure}
  \centering
  \includegraphics[width=8cm]{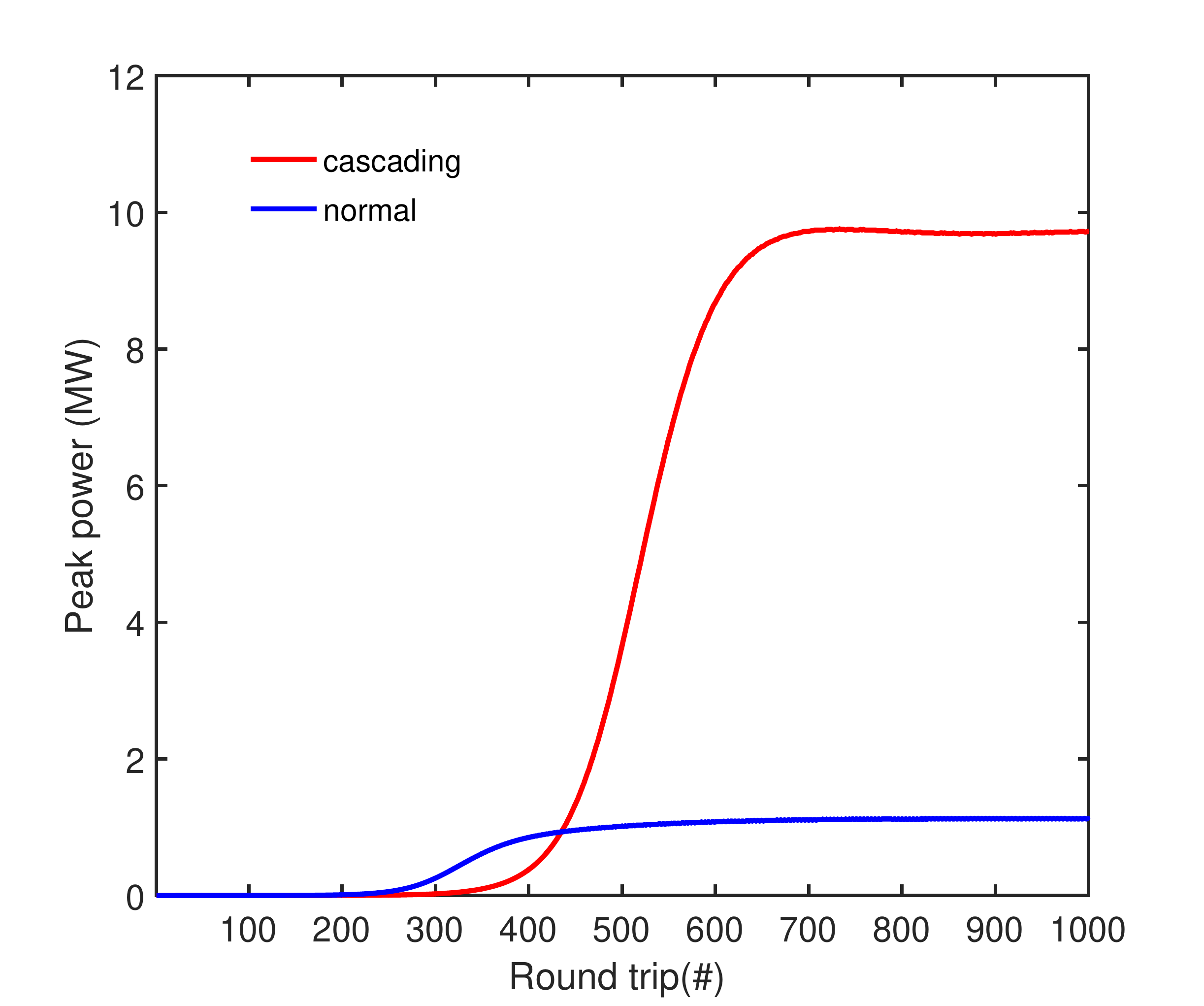}\\
  \caption{ (color online). The growth of output peak power in X-ray FEL oscillators from steady-state simulations, in which the total reflectivity and single-pass gain are assumed to be 89.8\% and 15\% respectively, and the output coupling is 4.5\%.}\label{x_round_trip}
\end{figure}

The results of gain cascading XFELO with three identical chicane delays of 500 fs is presented in Fig.~\ref{x_profile}, which shows the temporal and spectral power profiles of the out-coupled X-ray pulse after saturation. The finally output laser pulses are quasi-Gaussian distributions instead of step function profile are due to cut-off of Bragg crystals in the spectral domain. The output spectrum has a measured FWHM width of 9.14 meV, while the temporal FWHM width is about 460 fs. This corresponds to a bandwidth temporal product of 1, nearly twice the value of Fourier limit for a Gaussian pulse profile, which is due to the deviation of pulse profile from the Gaussian. According to Fig.~\ref{x_round_trip}, the peak power 9.7 MW is almost 8.6 times higher than 1.13 MW for the classical XFELO. It demonstrates 4.23 $\mu$J output pulse energy with nearly $1\times10^{-5}$, and comparing to the 1.67 $\mu$J given by the classical XFELO, the energy extraction efficiency increases by a factor of 2.5 which agrees reasonably well with 2.1 given by Eq.~(\ref{eq3}). Moreover, the energy spread of electron beam increases from 1.4 MeV to 2.25 MeV after FEL interaction which is still under control. The radiation pulse interaction with different (parts of) electron bunches inside each stage of undulator, thus the electron beam energy spread growth is assigned to different (parts of) electron bunches, which is good for the storage-ring based FEL oscillator \cite{45huang,44cai}.

The simulation results show that, the pulse energy loss due to the Bragg crystal reflection is large at the first a few round-trip, and drops to nearly $10^{-3}$ in the exponential growth region, which is due to the improvement of radiation longitudinal coherence. Although the energy loss is small for the whole pulse, it fluctuates rapidly at different parts of radiation because of the bandwidth stop of Bragg crystal. In practice, the heat loading effect is of vital importance and is analyzed in \cite{52song}. The part of light pulse which overlaps with the electron beam inside the four undulators gets the maximum net gain nearly 2.7\% and grows to saturation, while the other part of pulse does not obtain enough gain and cannot increase at all. In addition, the duration of the output pulse can be controlled by changing the length of the part of electron beam that overlaps with the radiation, and this can be simply achieved through the combine of different delay distance between the undulators. 

The robustness of our proposals is compared with normal FEL oscillator. For a particular wavelength, the requirements on the cavity tolerance and the electron beam parameters, e.g., the mirror alignment and beam energy spread in both schemes are similar in principle. Although optical pulses with much higher intensity are trapped inside the cavity, the beam energy spread induced by FEL interaction is under control because of undertaking of several stages. From the view point of flexibility, the proposed schemes are simple, and easy to implement. The simplest case only requires the installation of small chicane delays between certain undulator modules in addition to standard components. The proposed schemes could be easily tested on many of the present or future FEL oscillators.

\section{Conclusion}
In summary, we proposed a novel mode for FEL oscillators, in which the single-pass gain is from cascaded multi-stage undulators, and more accurately from different electrons. Compared with the performances of standard FEL oscillators, it is demonstrated with numerical simulations that the proposed schemes significantly improve the FEL oscillator performance: the radiation pulses peak power increase a factor of 2 for typical infrared oscillator; for an X-ray oscillator with four stages of undulators, the radiation peak power can be nearly 8.6 times larger with the energy extraction efficiency is 2.5 times higher. With the predicted characteristics, our schemes would contribute greatly to the FEL oscillators, and may open up new scientific opportunities in various research fields. It is worth stressing that the results here for the cascading FEL oscillator are not fully optimized, because a multi-dimensional scan of the performance, as a function of the cavity detuning and chicane delays is time-consuming procedures. Moreover, one may use different undulator and beam parameters for different stages, and then the proposals are naturally suited for producing simultaneous lasing at multiple wavelengths \cite{42wyk} or adjust the delay distance between the undulators to control the duration of output pulse.

\section{Acknowledgments}
The author would like to thank C. Feng, B. Liu, D. Wang, and Z. T. Zhao for enthusiastic discussions on FEL physics and beam dynamics. This work was partially supported by the National Natural Science Foundation of China (11322550 and 11475250) and Ten Thousand Talent Program.


\section{References}
\begin{thebibliography}{90}
\bibitem{1pc}Pellegrini C, Marinelli A and Reiche S 2016 {\it Rev. Mod. Phys} {\bf 88} 015006.
\bibitem{2elr}Elias L R, Fairbank W M and Madey J M J 1976 {\it Phys. Rev. Lett} {\bf 36} 717.
\bibitem{3vapw}Van Amersfoort P W, Bakker R J and Bekkers J B 1992 {\it Nucl. Instrum. Methods Phys. Res., Sect. A} {\bf 318} 42-46.
\bibitem{4ngr}Neil G R, Benson S V and Biallas G 2001 {\it Phys. Rev. Lett} {\bf 87} 084801.
\bibitem{5wyk}Wu Y K, Vinokurov N A and Mikhailov S 2006 {\it Phys. Rev. Lett} {\bf 96} 224801.
\bibitem{6gng}Gavrilov N G, Knyazev B A and Kolobanov E I 2007 {\it Nucl. Instrum. Methods Phys. Res., Sect. A} {\bf 575} 54-57.
\bibitem{7aw}Ackermann W, Asova G and Ayvazyan V 2007 {\it Nat. Photonics} {\bf 1} 336-342.
\bibitem{8ep}Emma P, Akre R and Arthur J 2010 {\it Nat. Photonics} {\bf 4} 641-647.
\bibitem{9it}Ishikawa T, Aoyagi H and Asaka T 2012 {\it Nat. Photonics} {\bf 6} 540-544.
\bibitem{10ae}Allaria E, Appio R and Badano L 2012 {\it Nat. Photonics} {\bf 6} 699-704.
\bibitem{10br}Bonifacio R, Pellegrini C, Narducci L M 1984 {\it Opt. Commun} {\bf 50} 373-378.
\bibitem{10cbr}Bonifacio R, De Salvo L, Pierini P 1994 {\it Phys. Rev. Lett} {\bf 73} 70.
\bibitem{11fa}Fratalocchi A and Ruocco G 2011 {\it Phys. Rev. Lett} {\bf 106} 105504.
\bibitem{12ss}Schreck S, Beye M and Sellberg J A 2014 {\it Phys. Rev. Lett} {\bf 113} 153002.
\bibitem{13sel}Saldin E L, Schneidmiller E A and Yurkov M V 2004 {\it Opt. Commun} {\bf 239} 161-172.
\bibitem{14zaa}Zholents A A and Fawley W M 2004 {\it Phys. Rev. Lett} {\bf 92} 224801.
\bibitem{15sel}Saldin E L, Schneidmiller E A and Yurkov M V 2006 {\it Phys. Rev. ST Accel. Beams} {\bf 9} 050702.
\bibitem{16xd}Xiang D, Huang Z and Stupakov G 2009 {\it Phys. Rev. ST Accel. Beams} {\bf 12} 060701.
\bibitem{17yj}Yan J, Deng H X and Wang D 2010 {\it Nucl. Instrum. Methods Phys. Res., Sect. A} {\bf 621} 97-104.
\bibitem{18ep}Emma P, Bane K and Cornacchia M 2004 {\it Phys. Rev. Lett} {\bf 92} 074801.
\bibitem{19rs}Reiche S, Musumeci P and Pellegrini C 2008 {\it Nucl. Instrum. Methods Phys. Res., Sect. A} {\bf 593} 45-48.
\bibitem{19atnr}Thompson N R, McNeil B W J 2008 {\it Phys. Rev. Lett} {\bf 100} 203901.
\bibitem{19bddj}Dunning D J, McNeil B W J, Thompson N R 2013 {\it Phys. Rev. Lett} {\bf 110} 104801.
\bibitem{19cmbw}McNeil B W J, Thompson N R 2012{EPL (Europhysics Letters)} {\bf 98} 29901.
\bibitem{20knm}Kroll N M, Morton P L and Rosenbluth M N 1979 {\it Phys. Quantum Electron} {\bf 7} 104.
\bibitem{21tt}Tanaka T 2013 {\it Phys. Rev. Lett} {\bf 110} 084801.
\bibitem{22pe}Prat E and Reiche S 2015 {\it Phys. Rev. Lett} {\bf 114} 244801.
\bibitem{23ylh}Yu L H 1991 {\it Phys. Rev. A} {\bf 44} 5178.
\bibitem{24sg}Stupakov G 2009 {\it Phys. Rev. Lett} {\bf 102} 074801.
\bibitem{25dh}Deng H and Feng C 2013 {\it Phys. Rev. Lett} {\bf 111} 084801.
\bibitem{26aj}Amann J, Berg W and Blank V 2012 {\it Nat. photonics} {\bf 6} 693-698.
\bibitem{27ht}Hara T, Inubushi Y and Katayama T 2013 {\it Nat. commun} {\bf 4} 2919.
\bibitem{28laa}Lutman A A, Decker F J and Arthur J 2014 {\it Phys. Rev. Lett} {\bf 113} 254801.
\bibitem{29ma}Marinelli A, Ratner D and Lutman A A 2015 {\it Nat. commun} {\bf 6} 6369.
\bibitem{30yj}Yan J, Hao H, Li J Y 2016 {\it  Phys. Rev. ST Accel. Beams} {\bf 19} 070701.
\bibitem{30hz}Huang Z, Kim K J 2007 {\it  Phys. Rev. ST Accel. Beams} {\bf 10} 034801.
\bibitem{30qb}Qin B, Tan P, Yang L 2013 {\it Nucl. Instrum. Methods Phys. Res., Sect. A} {\bf 727} 90-96.
\bibitem{30lk}Li K, Song M and Deng H 2017 {\it  Phys. Rev. ST Accel. Beams} {\bf 20} 030702.
\bibitem{31mp}Musumeci P, Moody J T and Scoby C M 2010 {\it  J. Appl. Phys} {\bf 108} 114513.
\bibitem{32ff}Fu F, Wang R and Zhu P 2015 {\it Phys. Rev. Lett} {\bf 114} 114801.
\bibitem{33sy}Shen Y, Yang X and Carr G L 2011{\it Phys. Rev. Lett} {\bf 107} 204801.
\bibitem{34zz}Zhang Z, Yan L and Du Y 2016 {\it Phys. Rev. Lett} {\bf 116} 184801.
\bibitem{34wj}Wu J and Yu L H 2001 {\it Nucl. Instrum. Methods Phys. Res., Sect. A} {\bf 475} 104-111.
\bibitem{35kkj}Kim K J, Shvyd¡¯ko Y and Reiche S 2008 {\it Phys. Rev. Lett} {\bf 100} 244802.
\bibitem{36lrr}Lindberg R R, Kim K J, Shvyd¡¯ko Y 2011 {\it  Phys. Rev. ST Accel. Beams} {\bf 14} 010701.
\bibitem{37dj}Dai J, Deng H and Dai Z 2012 {\it Phys. Rev. Lett} {\bf 108} 034802.
\bibitem{38vna}Vinokurov N A and Skrinsky A N 1977 {\it preprint INP77} {\bf 59}.
\bibitem{39dib}Drobyazko I B, Kulipanov G N and Litvinenko V N 1989 {\it Nucl. Instrum. Methods Phys. Res., Sect. A} {\bf 282} 424-430.
\bibitem{40pjm}van der Slot P J M, Freund H P and Miner Jr W H 2009 {\it Phys. Rev. Lett} {\bf 102} 244802.
\bibitem{41rs}Reiche S. 1999 {\it Nucl. Instrum. Methods Phys. Res., Sect. A} {\bf 429} 243-248.
\bibitem{41ngr}Neil G R, Behre C, Benson S V 2006 {\it Nucl. Instrum. Methods Phys. Res., Sect. A} {\bf 557} 9-15.
\bibitem{42JD}Jaroszynski D A, Chaix P, Piovella N, 1997 {\it Nucl. Instrum. Methods Phys. Res., Sect. A} {\bf 393} 332-338.
\bibitem{44cai}Cai Y, Ding Y, Hettel R 2013 {\it Synchrotron Radiat. News} {\bf 26} 39-41.
\bibitem{45huang}Huang Z, Bane K, Cai Y 2008 {\it Nucl. Instrum. Methods Phys. Res., Sect. A} {\bf 593} 120-124.
\bibitem{52song}Song M Q, Zhang Q M, Guo Y H 2016 {\it Chinese Phys. C} {\bf 40} 048101.
\bibitem{42wyk}Wu Y K, Yan J and Hao H 2015 {\it Phys. Rev. Lett} {\bf 115} 184801.
\end{thebibliography}
\end{document}